# Inverse Problems in Musical Instrument Modeling: A Structured Taxonomy and Review

--Manuscript Draft--

| | |
|---|---|
| **Manuscript Number:** | |
| **Full Title:** | Inverse Problems in Musical Instrument Modeling: A Structured Taxonomy and Review |
| **Article Type:** | ISMA26 |
| **Corresponding Author:** | Xinmeng Luan<br>McGill University<br>Montréal, Quebec CANADA |
| **Order of Authors:** | Xinmeng Luan |
| | Gary Scavone |
| **Abstract:** | Inverse problems arise in a wide range of applications in musical acoustics, including physics-based sound synthesis, musical instrument modeling, design, and optimization. However, these problems are inherently challenging due to their ill-conditioned nature and strong sensitivity to measurement noise. This paper presents a structured taxonomy and systematic review of inverse problems in musical instrument modeling, providing a unified framework for their classification and analysis. We categorize inverse problems in musical instrument modeling into ten distinct tasks: mechanical parameter estimation; geometric parameter estimation; loss estimation; boundary condition estimation; modal parameter estimation; excitation and articulatory parameter estimation; sound matching or model parameter fitting; instrument design and optimization; field reconstruction, characterization and separation; physical model identification and discovery. |
| **Section/Category:** | Musical Acoustics |
| **Additional Information:** | |
| **Question** | **Response** |
| Please read the Instructions for full information. Select the appropriate information below to verify that you have consented to this copyright licensing and the conditions and representations set forth in the Instructions. |  |

## 1. INTRODUCTION

Musical instrument modeling[1] has become a central topic in musical acoustics, aiming to understand, reproduce, and synthesize sound generation mechanisms through physical and computational approaches, as well as to support instrument design and optimization. In particular, physics-based sound synthesis has gained increasing attention, as it enables both realistic audio generation and deeper insight into the underlying vibroacoustic behavior of instruments. It also provides a framework for optimizing model parameters to improve sound quality and instrument design. Within this context, inverse problems play a key role by offering a systematic approach to infer physical parameters or hidden variables from measured acoustic data.

In general, inverse problems in musical acoustics consist in estimating unknown quantities of a physical system from observations such as sound or vibration signals. Typical for inverse problems, a major challenge lies in the inversion of the transfer or propagation operator describing the forward physical process.[2] This operator is often ill-conditioned, either due to the physics of wave propagation or numerical discretization, making its inversion highly sensitive to small perturbations. In practice, measurement noise is unavoidable and can lead to large errors in the reconstructed solution. Therefore, regularization techniques are essential to stabilize the inversion and reduce noise amplification.

Within this context, we categorize inverse problems in musical instrument modeling into ten tasks. To maintain a consistent and general framework, we formulate most of these tasks within this paradigm. It should also be noted that some problems can be expressed in multiple ways and may naturally fall into more than one category depending on the chosen formulation. Furthermore, the examples provided for each task tend to be focused mainly on wind or string instruments and do not represent a comprehensive list.

## 2. INVERSE PROBLEMS

### A. MECHANICAL PARAMETER ESTIMATION

Mechanical parameters of musical instruments are sometimes difficult to measure directly, motivating the use of inverse methods to infer them from observed data.

Representative studies span multiple instrument families. For woodwinds, research has focused on estimating reed parameters in clarinets under real playing conditions using a two-stage optimization strategy,[3] as well as on saxophone reed parameter estimation during performance.[4] In brass instruments, efforts have been made to identify lip-reed parameters within physical modeling frameworks.[5,6] For string instruments, inverse approaches have been used to estimate the elastic properties of spruce plates[7] and to characterize tonewood materials for instrument soundboards.[8] Additionally, methods based on least-squares fitting and Chladni patterns have been proposed to estimate the elastic constants of rectangular orthotropic plates.[9] Bow-string mechanical parameters in an elasto-plastic friction model were estimated through inverse modeling to reconstruct transient responses.[10]

### B. GEOMETRY PARAMETER ESTIMATION

While wooden instruments can generally be imaged using X-ray CT, metallic instruments often require neutron CT, which is less accessible, making the acquisition of high-quality scans of brass instruments, flutes, and saxophones particularly challenging.[11] Bore reconstruction and geometry estimation have long been important topics in wind instrument research.

For simple wind instruments without side holes (e.g., brass instruments or isolated woodwind sections), early studies mainly used acoustic pulse reflectometry to reconstruct geometries such as trumpets[12] and bassoon crooks.[13,14] Zeroth-order optimization methods, including the Rosenbrock algorithm, have also been applied to trumpet geometry reconstruction,[15] while gradient-based approaches such as the Levenberg-Marquardt algorithm have been used for bore reconstruction with segmented conical or cylindrical models.[16] For instruments with tone holes, the problem becomes more complex due to multipath propagation and coupling effects. In this case, pulse reflectometry can lead to errors, as wave propagation in chimneys distorts the effective upstream bore, particularly when holes are closed.[17] More advanced techniques, such as full waveform inversion, have been proposed to jointly recover bore geometry and tone-hole configurations in woodwind instruments.[18]

### C. LOSS ESTIMATION

Dissipative phenomena are inherent to all musical instruments, as part of the input mechanical energy is converted into heat or radiated as sound.[1] Modal methods offer a key advantage for damping modeling: losses can be assigned and tuned independently for each mode with minimal computational cost. This enables direct use of measured loss parameters, effectively turning modal synthesis into an analysis-synthesis framework. Consequently, individual partials can be reproduced with high accuracy.[19] Loss estimation in modal form is further discussed in Sec. 2.E.

Virtual wind instrument prototyping relies on accurate modeling of sound propagation in ducts with varying cross-sections, typically described by transmission-line equations. To account for viscous and thermal boundary layer effects at the tube walls, the coefficients of these models incorporate the fluid's viscosity and thermal properties. A classical formulation for cylindrical tubes was introduced by Zwikker and Kosten.[20] This approach was later generalized to straight tubes with arbitrary cross-sectional shapes,[21] and more recently extended to derive transmission line coefficients for viscothermal acoustics in conical tubes.[22]

Two other illustrative examples include the prediction of decay times in solid-body electric guitar tones,[23] where both intrinsic string damping and damping arising from mechanical coupling with the guitar neck are identified. Another example involves the inference of drumhead damping and tuning parameters from Finite-Difference Time-Domain (FDTD)-simulated sounds using a convolutional neural network (CNN).[24]

### D. BOUNDARY CONDITION ESTIMATION

Boundary conditions play a crucial role in determining the dynamic behavior of musical instruments, as they govern how energy is injected, transmitted, and radiated within the system.

Regarding acoustic tube radiation modeling, the geometry of the opening significantly influences the resulting radiation behavior. Numerical approximations of the radiation impedance for tubes with different flange configurations have been investigated in Dalmont et al.[25] More recently, physics-informed neural network (PINN) approaches have been explored for estimating radiation parameters in acoustic tubes.[26] In practice, however, radiation conditions are often more complex due to irregular geometries at the tube termination, as encountered in wind instruments, for example, configurations with multiple openings distributed along the main bore.[27] Additionally, external tonehole interactions, including mutual radiation effects between openings, play a significant role in woodwind instruments and must be considered for accurate modeling.[28]

The surface velocity of a structure, interpreted as a boundary condition for the external radiation field, can be reconstructed using Near-field Acoustic Holography (NAH) from near-field pressure measurements acquired by microphone arrays. Compared to Laser Doppler Vibrometry (LDV), NAH is a more cost-effective alternative in terms of instrumentation, although its accuracy strongly depends on the chosen inverse reconstruction algorithm. NAH has been applied, for instance, to the study of the Couchet harpsichord soundboard[29] and the violin.[30–32] More recently, physics-informed deep learning methods have been introduced for NAH and have been applied to the reconstruction of violin top plate vibrations.[33–36]

## E. MODAL PARAMETER ESTIMATION

As a common post-processing step, modal parameter estimation is applied to measured admittance or impedance curves, for instance, to analyze and simulate the bridge admittance of a string instrument or the input impedance of a wind instrument.

For wind instruments, modal parameters are typically obtained by locally approximating each resonance.[37] A common approach follows a two-step procedure,[38] where system poles are first identified and subsequently used to compute modal residues.[39,40] A similar strategy is employed in Taillard et al.,[41] combining pole estimation with a subsequent parameter fitting stage. ESPRIT[42] is particularly effective in cases of strong modal overlap and have been widely applied across different instrument families.[43–45] Alternative formulations model the input impedance as a recursive digital filter with frequency-dependent pole allocation.[46] More recently, hybrid approaches have been proposed in which multiple modes are allowed to contribute in the vicinity of each resonance.[47] In addition, a direct method for computing modal coefficients of the input impedance from the known instrument geometry has been introduced in Chabassier et al.[48]

For string instruments, spectrogram-based analysis methods introduced in[49] have been used to estimate eigenfrequencies and damping characteristics.[50–52] ESPRIT[42] has been applied to extract modal parameters of strings.[23,53,54] Filter Diagonalisation Method (FDM)[55] has been employed to extract the modal paramters of cello body impulse response.[56,57] In addition, a comparative study of four modal parameter extraction methods for string instruments is presented in Rau et al.[58]

## F. EXCITATION AND ARTICULATORY PARAMETER ESTIMATION

Excitation mechanisms in musical instruments are highly nonlinear and involve complex human interactions such as bowing, fingering, and breath control. End-to-end Music Information Retrieval (MIR) approaches, which learn mappings directly from data without incorporating physical constraints, are outside the scope of this paper.

For example, the force at the bow-string interaction is difficult to measure directly in experiments. To address this challenge, Schumacher[59] proposed a method to reconstruct the bowing force from force signals measured at the string terminations. Subsequently, two reconstruction approaches-one in the time domain and the other in the frequency domain-were developed. When applicable, both methods yield consistent and reliable results.[60] In addition, a modal-based identification approach based on inverse methods was developed to estimate the friction force and string dynamics at the bow-string contact point using end-support reaction measurements.[61] Moreover, a method for estimating the plucking point and pickup position of an electric guitar based on the autocorrelation of spectral peaks is proposed in Mohamad et al.[62]

## G. SOUND MATCHING OR MODEL PARAMETER FITTING

Sound matching,[63] or model parameter fitting, aims to adjust the parameters of a sound synthesis model so that its output closely matches measured audio signals. This step is crucial because computational models are inherently approximations of real instruments, and discrepancies inevitably arise, for example due to neglected coupling effects or simplified physical assumptions. Typical applications include estimating Digital Waveguide (DWG) filter coefficients to reproduce recorded sounds, or tuning numerical models to match measured resonances and modal behavior. Sound matching primarily targets perceptual similarity, often assessed using signal-based perceptual metrics, and does not necessarily guarantee physically meaningful parameter estimates. As a result, recovering physically consistent parameters is significantly more challenging than achieving perceptual sound matching.

Cemgil et al.[64] formulates physical model parameter estimation of plucked strings as learning an inverse mapping from signal features to model parameters via a multilayer perceptron (MLP). A scattering recurrent network (SRN) was proposed to simulate plucked string vibration, with system parameters learned via

BackPropagation Through Time (BPTT).[65,66] A signal-analysis approach combined with derivative-free optimization was proposed for guitar string simulation in a DWG framework.[67] A parameter estimation method using a genetic algorithm and a perceptual fitness function was developed and evaluated for a plucked-string DWG model.[68] Parameter estimation for piano sound synthesis based on physical modeling was performed using a FDTD approach combined with gradient-based optimization.[69] A multi-stage end-to-end CNN framework was proposed to estimate DWG coefficients of a pipe organ directly from audio signals.[70] In addition, a deep learning-based perceptual sound matching approach was integrated within a Functional Transformation Method (FTM) framework for drum sound synthesis.[63] Inverse modeling was performed by directly optimizing physical parameters of strings, membranes, and plates, rather than conventional spectral parameters such as poles and zeros, within a differentiable modal framework.[71]

## H. INSTRUMENT DESIGN AND OPTIMIZATION

Inverse modeling offers a powerful framework for optimizing both the geometry and physical properties of musical instruments. In these approaches, a desired target acoustic response is specified, and design parameters are iteratively adjusted to achieve it, typically using numerical simulations or surrogate models. Such methods can significantly benefit luthiers and instrument makers by providing a more systematic, scientifically grounded workflow for instrument design and refinement.

For wind instruments, a wide range of optimization strategies has been investigated for instrument design. Noreland[72] used a gradient-based method with analytic derivatives to design brass bore profiles composed of truncated cones. Zeroth-order optimization techniques, such as the Rosenbrock algorithm, have also been applied to trombone bore design.[73] Colinot et al.[74] designed a coaxial saxophone without side holes, effectively equivalent to a conical instrument. For tone-hole instruments, early work by Debut et al.[75] focused on adjusting register hole placement and main bore parameters of a clarinet. Subsequent studies employed gradient-based optimizers such as L-BFGS-B to jointly optimize tone-hole locations, radii, and bore length.[76] A related approach led to the concept of a "logical clarinet," in which each note is produced by a dedicated hole without fork fingering, with optimized hole geometry while keeping the main bore fixed.[77] Ernoult et al.[78] further explored the optimization of woodwind instruments with non-cylindrical bores and detailed tone-hole structures using a sequential quadratic programming (SQP) method. In a similar spirit, Tournemenne et al.[79] proposed a sound simulation-based optimization framework for the design of brass instrument resonator geometries.

As for string instruments, a model for sound tuning in asymmetrically braced guitars was proposed.[80] In addition, geometrically guided shaping of guitar soundboards has also been investigated.[81]

## I. FIELD RECONSTRUCTION, CHARACTERIZATION AND SEPARATION

Inverse methods provide a means of reconstructing high-resolution physical fields from sparse measurements. A closely related application is the characterization of musical instrument radiation. The radiation directivity is typically measured using sparse microphone arrays, requiring interpolation and extrapolation to recover the full spatial distribution. Spherical harmonics are commonly used to represent instrument directivity from measured data, providing a compact and mathematically well-founded basis for interpolation.[82] Alternatively, the Helmholtz Equation Least Squares (HELS) method formulates directivity estimation as a spherical wave decomposition of the radiated field.[83] More recently, physics-informed approaches have been introduced to improve interpolation and extrapolation performance by embedding physical constraints into the learning process.[84,85] Moreover, the NAH problem discussed in Sec. 2.D can also be viewed as a vibrational field reconstruction problem.

For wind instruments, characterization of acoustic field in waveguide through measurement often relies on wave separation techniques. A classical two-microphone method enables the decomposition of left-and right-traveling waves inside instruments such as the trombone, allowing the identification of nonlinear effects associated with progressive wave propagation.[86] Building on this idea, time-domain approaches have been proposed in which propagation losses are explicitly accounted for, with successful application to trumpets under playing conditions.[87, 88] Adaptive methods have been developed to estimate acoustic variables in ducts under time-varying propagation conditions, which, under the plane-wave assumption, can be formulated as a traveling wave separation problem.[89] Recently, PINN approaches have been explored for one-dimensional time-domain acoustic field reconstruction in tubes.[26]

### J. PHYSICAL MODEL IDENTIFICATION AND DISCOVERY

Another class of inverse problems aims to discover or refine the physical models governing instrument behavior, particularly when the underlying mechanisms are not fully understood or when simplifying numerical assumptions are required. In such cases, the objective is often to identify an appropriate functional representation of the excitation, interaction, or wave propagation mechanisms, informed by experimental measurements and numerical simulations. The emphasis in this section is therefore on the model structure itself, rather than on parameter estimation within a predefined model.

Numerous studies have focused on determining the series and shunt impedances of tonehole models, which provide a simplified 1D equivalent representation of inherently 3D geometries.[90–94] A prominent example is the bow-string interaction, which remains an active area of ongoing research and model development. In this context, inverse modeling is used to estimate parameters of elastoplastic friction models, enabling accurate reconstruction of measured transient signals.[95]

## 3. CONCLUSION

This paper presents a structured taxonomy and review of inverse problems in musical instrument modeling, categorizing them into ten distinct tasks. Despite their diverse objectives and formulations, these problems share common challenges, particularly ill-conditioning and sensitivity to measurement noise. The proposed taxonomy provides a unified framework for organizing existing approaches while highlighting connections both within and across different inverse problems.